\documentclass[prl,twocolumn,floatfix,showpacs,superscriptaddress]{revtex4-1} 

\usepackage{amsmath}

\usepackage{color, amssymb, graphicx, amsfonts,epstopdf }

\newcommand{\ket}[1]{\ensuremath{\left|#1\right\rangle}}

\begin{document}
\title{An unusual superconducting Isotope Effect in the presence of a Quantum Criticality}

\author{Yaron Kedem}
\affiliation{Nordita, Center for Quantum Materials, KTH Royal Institute of Technology and Stockholm University, Roslagstullsbacken 23, 10691 Stockholm, Sweden}
\author{Jian-Xin Zhu}
\affiliation{Theoretical Division, Los Alamos National Laboratory, Los Alamos, New Mexico 87545}
\affiliation{Center for Integrated Nanotechnologies, Los Alamos National Laboratory, Los Alamos, New Mexico 87545}
\author{Alexander V. Balatsky}
\affiliation{Institute for Materials Science, Los Alamos National Laboratory, Los Alamos,
NM, 87545, USA}
\affiliation{Nordita, Center for Quantum Materials, KTH Royal Institute of Technology and Stockholm University, Roslagstullsbacken 23, 10691 Stockholm, Sweden}

\date{\today}
\begin{abstract}
We discuss the possible connection between superconductivity (SC) and quantum critical points (QCP) for any QCP that is tunable by isotopic mass substitution. We find a distinct contribution to the isotope exponent, due to the proximity to a QCP, which can be used as an experimental signature for the relation between SC and QCP. The relation is demonstrated in a scenario where the SC pairing is due to modes related to a structural instability. Within this model the isotope exponent is derived in terms of microscopic parameters.     
\end{abstract}

\maketitle

The explanation of the isotope effect \cite{isotope1,isotope2} is one of the most celebrated triumphs of the BCS theory\cite{bcs}. Its usual form, where the transition temperature is inversely proportional to the square root of the isotopic mass, highlights both the role of phonons in the pairing mechanism and the validity of the main BCS assumption, namely a constant attractive interaction between electrons within the Debye energy. In superconductors which cannot be described by the BCS theory, such as High-Tc superconductors, strong deviations from the simple BCS form are observed\cite{frank}. In general, a transition temperature that is independent on the isotopic mass is regarded as evidence against the role of phonons in the pairing mechanism \cite{NoIso1,NoIso2}. Further studies revealed the effect of isotopic replacement on many other quantities, such as the pseudogap \cite{Pseudo}, magnetic penetration depth \cite{penetDepth}, electron dynamics \cite{arpes} and quasiparticle inelastic scattering \cite{lee2006interplay} . This evidence motivated a variety of theoretical ideas for the pairing mechanism, such as the bipolaron theory \cite{bipolaron}, magnetically mediated SC \cite{magSC1,magSC2} and valence fluctuation \cite{valence}.

The notion that superconductivity is facilitated by a proximity to a quantum critical point (QCP) has been widely discussed \cite{Gegenwart2008,Sachdev2000,zaanen}. In the literature, it is common to focus on the competing phases existing close to a QCP and to infer that due to this competition, an opening emerges for new phases to appear, of which superconductivity may be one. Thus the QCP is considered an established mechanism for the formation of a superconducting dome. Several types of quantum phases were proposed for the relevant QCP, for example magnetic order \cite{magSC1}, charge order \cite{CDWsc} or metal insulator transition \cite{MetalInsulatorSC}. Changing one physical parameter, such as doping level or pressure, leads to changes of multiple interactions. Therefore the challenges in discussions on QCP mechanism is to disentangle the relative importance of various effects on superconductivity. Finding a distinct experimental signature, connecting the transition temperature to the proximity to a QCP, remains a theoretical challenge. 

\begin{figure}
\centering
\includegraphics[trim=7.7cm 3.05cm 11.8cm 4cm, clip=true, width=\columnwidth]{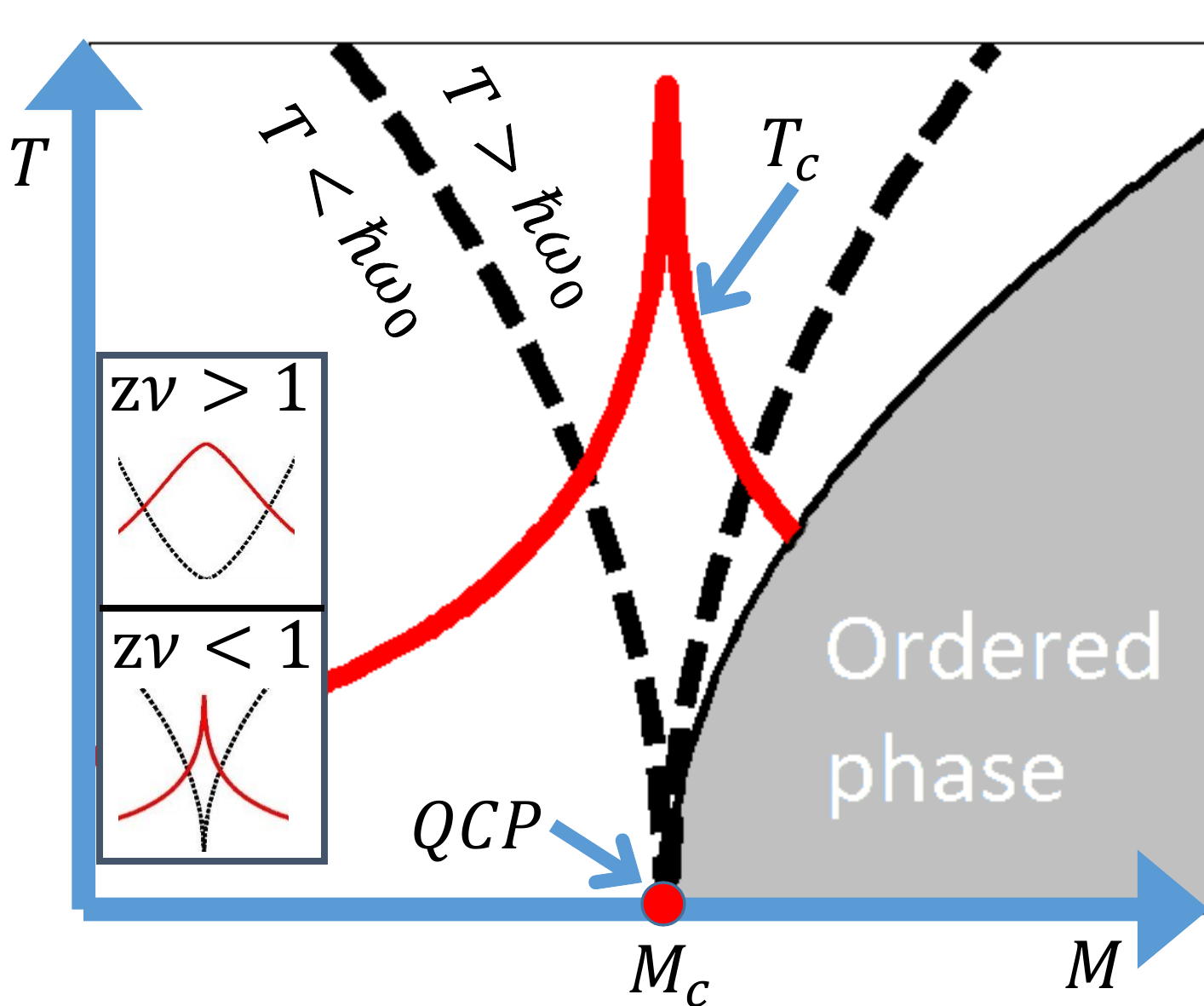}
\caption{An illustration of the dependency of $T_c$ on the isotopic mass $M$, in the vicinity of a QCP. There is a structural phase transition to an ordered, or symmetry breaking, phase (shown as a shaded area on the bottom right corner). The transition temperature to this phase depends on the isotopic mass $T(M)$ and defines the edge of the shaded area. At the QCP, where $M=M_c$ this temperature vanishes $T(M_c)=0$ so the system cannot reach the ordered phase for lower values of $M$. The dashed lines represents a crossover to the quantum critical region in which $\hbar \omega_0 < T$, where $\omega_0$ is a frequency of a structural mode related to the phase transition and it vanish at the QCP as shown in (\ref{w0}). Assuming this mode is responsible for superconductivity, we also plot the superconducting transition temperature in the form $T_c \sim e^{- \omega_0 / a}$ (solid line). Since the crossover line is given by $T = \hbar \omega_0$, it is a mirror image of  $T_c \sim 1 - \omega_0 $. At the QCP, where $ \omega_0  \sim |M-M_C|^{z\nu}$, these lines can have a cusp, with diverging derivative, or be smooth, with vanishing derivative, depending whether $ z\nu > 1$ or $ z\nu < 1$, as illustrated in the inset.}
\label{qcp}
\end{figure}

Recently a QCP related to a ferroelectric order was used to explain the enigmatic superconducting dome in strontium titanate (STO)\cite{sto}. Remarkably, this order can be tuned via isotopic replacement and the QCP is reached at 35\% replacement of O$^{16}$ to O$^{18}$ in STO \cite{Itoh1999,rowley}. This fact allows the use of the isotope effect to investigate the connection between a QCP and superconductivity. Here, we expand the approach used in \cite{sto} in order to demonstrate how a clear signature of quantum criticality can be shown via the isotope effect. Employing a simple model for the transition temperature, we identify different contributions to the isotope exponent and show in which regime the contribution due to the quantum criticality is dominated. 

We start by deriving a general expression for the isotope exponent, using a simple form for $T_c$ and the coupling constant. Later, we examine a specific case where this expression is applicable. We consider a QCP, which is due to a structural instability, where the structural modes, related to the instability, are responsible for the electron-electron coupling.     

A general expression for $T_c$ can be written as
\begin{equation} \label{tc}
T_c = \Theta f(\lambda) 
\end{equation}
where $\Theta$ is the energy scale and $ f(\lambda)$ is a dimensionless function of the dimensionless coupling constant $\lambda$. The isotope exponent is defined as $\alpha_T = -\left( {M \over T_c} \right) { \partial T_c \over \partial M}$, where $M$ is the isotopic mass. Using Eq. (\ref{tc}) we get two separate contributions. The first one is due to the energy scale
\begin{equation} \label{alphT}
\alpha_\Theta = - \left( {M \over \Theta} \right) { \partial \Theta \over \partial M}. 
\end{equation}
In the usual BCS scenario where the scale is given by the Debye temperature $\Theta = T_D \sim M^{-1/2}$ and $ { \partial \lambda \over \partial M}=0$, we have $\alpha_T = \alpha_\Theta = 1/2$. The additional contribution is given by 
\begin{equation} \label{extra}
 - \left( {M \over f(\lambda)} \right) { \partial f(\lambda) \over \partial \lambda} { \partial \lambda \over \partial M}.
\end{equation}
In order to calculate this contribution, let us consider a simple form for $ f(\lambda)$ and $\lambda$. The function $ f(\lambda)$ usually arises from a solution of a self consistent equation for $T_C$. In the so-called logarithmic approximation, this solution is given by the form $ f(\lambda) = e^{-1/\lambda}$ \cite{gorkov1,AGD}. The coupling constant represents an effective attractive interaction between electrons and we assume that this is due to a soft mode related to the QCP. So we use the form $\lambda \simeq a / \omega_0$, where $\omega_0$ is the frequency of the soft mode and $a$ is a factor incorporating the coupling strength between electrons and the soft mode. This form can be derived using the McMillan formula \cite{McMillan1968}
\begin{equation} \label{lam}
\lambda=\int_ 0^{\infty} \alpha^2(\omega) F(\omega) {d\omega \over \omega},
\end{equation}
where $\alpha(\omega)$ (not to be confused with the exponent $\alpha_T$, $\alpha_\Theta$ etc.) is the electron-phonon coupling and $F(\omega)$ is the spectral density of the phonons. The integral in Eq. (\ref{lam}) is typically dominated by the lowest frequency and thus we approximate $\lambda$ as $\lambda \simeq a / \omega_0$. The approximation becomes particularly good when the spectrum has a van Hove singularity, but it captures the main features of Eq. (\ref{lam}) for a wide variety of systems.

At the QCP, which is tunable via $M$, the frequency of the soft mode should vanish so we can write it as  \cite{sachdevBook}
\begin{equation} \label{w0}
\omega_0= \omega_s \left|{M-M_c \over M}\right|^{z \nu},
\end{equation}
where $M_c$ is the mass at the critical point, $\omega_s$ is an energy scale and $z \nu$ is the critical exponent of the system. We follow the conventional notation where $\nu$ is the exponent of the correlation length and $z$ is the dynamical factor, even though in this work $z \nu$ can be regarded as a single quantity. The behavior of the soft mode near the QCP, shown in Eq. (\ref{w0}), is crucial for our result. Below, we derive it for a specific example but essentially, it comes from the nature of a quantum phase transition. The justification for using $M$ as the critical parameter comes from the experiments  \cite{Itoh1999,rowley} which observed such transition via isotope replacement.

Using these expressions for $\lambda$ and $\omega_0$ in (\ref{extra}), we get two additional contributions to the critical exponent such that $\alpha_T = \alpha_\Theta + \alpha_{AH} + \alpha_c $. The first one
\begin{equation} \label{alph2}
\alpha_{AH} =  M \left|{M-M_c \over M}\right|^{z \nu} { \partial \over \partial M} \left({\omega_s \over a} \right)
\end{equation}
is due to anharmonicity of the mode, since ${ \partial \over \partial M} \left({\omega_s \over a} \right)=0$ for harmonic modes  \footnote{For harmonic modes, the displacement operator $X$ is proportional to $M^{-1/4}$ and $\omega_s \sim M^{-1/2}$. Since the electron-phonon coupling strength, $a$ is proportional to the square of the amplitude of the displacement $a \sim X^2 \sim M^{-1/2}$, the ratio ${\omega_s \over a}$ is, in general, mass independent.}. The second one  
\begin{equation} \label{alphc}
\alpha_{c} = \text{sgn}(M-M_c) \left( {M_c \over M}\right) \left({\omega_s \over a}\right) z \nu \left|{M-M_c \over M}\right|^{z \nu-1} 
\end{equation}
is due to the critical behavior of the mode. The sign in (\ref{alphc}) is determined by the side of the QCP on which the system lies, i.e. whether increasing the mass moves the system closer to the QCP or further away.

The dependence of $T_c$ on $M$ around the QCP is illustrated in Fig. \ref{qcp}. As the system approaches the QCP, $\alpha_\Theta$ is largely unaffected and $\alpha_{AH}$ is decreasing. The behavior of $\alpha_{c} $ depends on whether $z \nu$ is smaller or bigger than unity. For $z \nu < 1$, the critical contribution $\alpha_{c} $, given by Eq. (\ref{alphc}), will dominate the isotope exponent. This result is general and applicable to any superconductor in which the pairing is due to a soft mode near a QCP that can be tuned via the isotopic mass. 

For $ M < M_c$ the isotope exponent will take negative values, meaning an enhancement of $T_c$ with an increase in the mass. Such a behavior is regarded as anomalous, compared to the usual BCS result of $\alpha = 1/2$. The result shown by Eq. (\ref{alphc}) is a clear experimental signature for the contribution of a QCP to superconductivity. This method is accessible for a wide range of systems. In order to access the quantum critical regime it might be necessary to change $M$ continuously, which can be done by  changing the fractional composition between two isotopes, thereby continuously changing the effective isotope masses. This is because criticality usually arises from collective modes.   

The connection between the superconducting coupling constant $\lambda$ and a soft mode with vanishing frequency at the QCP can be valid in many systems. It can be instructive to consider a concrete example, where one can calculate $\omega_0$ to obtain Eq. (\ref{w0}). To this end, we would like to study phonons that are related to a structural instability which can potentially lead to breaking of a lattice symmetry. At zero temperature the lattice still remains in the high symmetry state, due to the quantum uncertainty in the position of the ions. Tuning some parameter, in our case the mass of the ions, can tip the system into the broken symmetry state once the QCP is reached. The phonons we consider are the modes which break the symmetry of the lattice.

\begin{figure}
\centering
\includegraphics[width=\columnwidth]{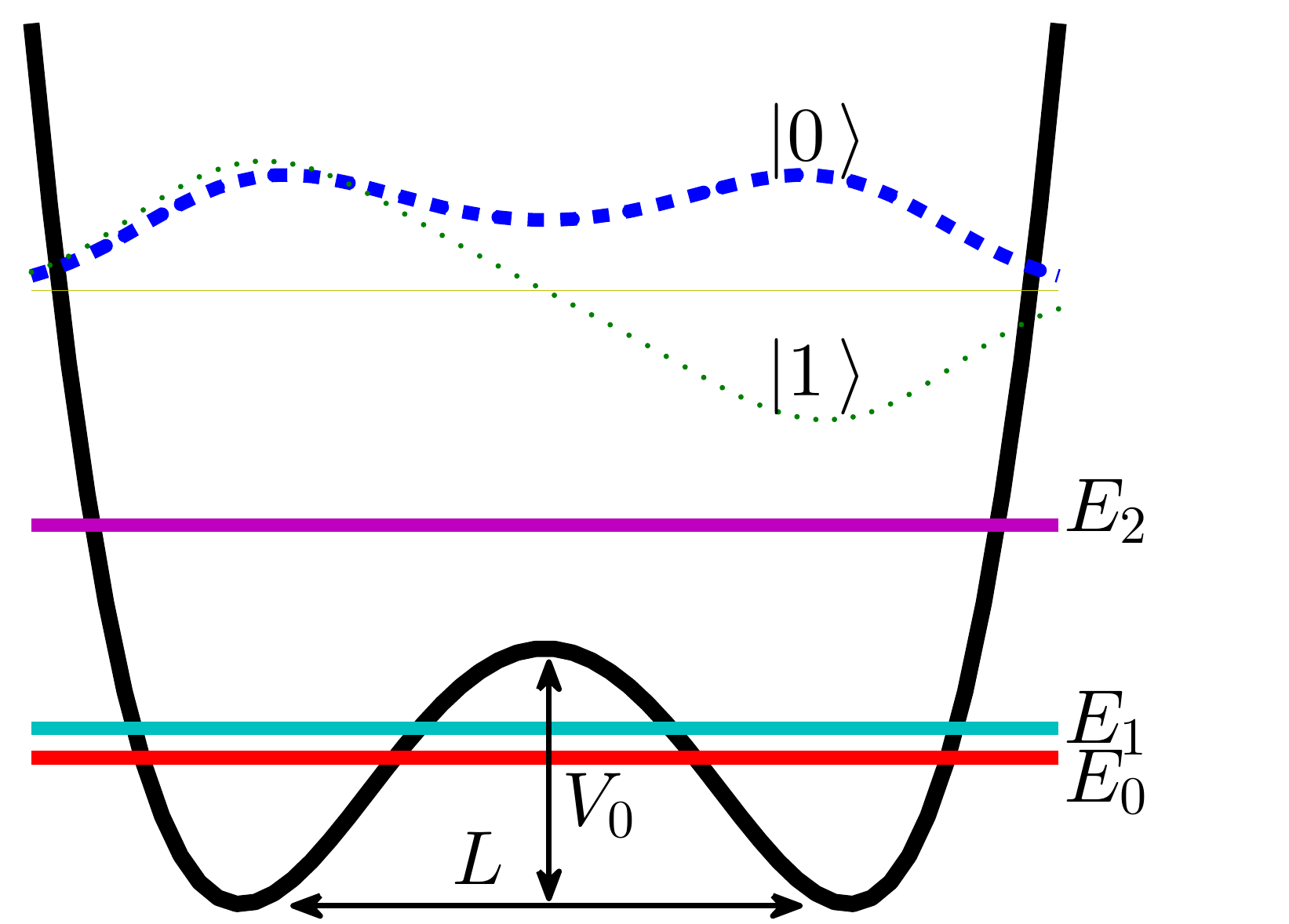}
\caption{An illustration of a system described by (\ref{h1}), with a potential of the form $V(A) = V_0 \left[ \left(A \over L\right)^4 - 2 \left( A \over L\right) ^2 \right]$. The double well potential is shown in scale with the lowest 3 energy levels. For the chosen parameter regime, $M=1$, $L=1$, $V_0 = 4$ and $\hbar=1$, the first 2 levels are much closer so higher levels can be ignored. The wave functions of these levels are also shown, where the (anti) symmetric ones is the (first excited) ground  state.
}
\label{cartoon}
\end{figure}

Consider a structural mode $\vec{e}_i$ of a single unit cell, i.e. an optical phonon, pertaining to a structural instability. The displacement of the ion $i$ in the unit cell, relative to the high symmetry state, is given by $\vec{r}_i =A \vec{e}_i $, where $A$ is the amplitude of the mode. The amplitude $A$ represents an effective coordinate on which we develop a low energy theory. The kinetic energy is given by $T= {1 \over 2} \sum_i m_i \left|\dot{\vec{r}_i} \right|^2 = {1 \over 2} M \dot{A}^2$, where $m_i$ is the mass of the ion and
\begin{equation} \label{M}
M= \sum_i m_i \left(\vec{e}_i \right)^2.
\end{equation}
is the effective mass of the mode. This sets a concrete relation between the mass of the atoms, which are affected by isotope replacement, and the effective mass $M$ so we treat the latter as the isotopic mass. The instability of the mode implies that the potential energy of the system, as a function of $A$, is given by a double well form, i.e. the minima exist at $A \ne 0$ when the symmetry is broken. This is illustrated in Fig. \ref{cartoon}.

\begin{figure}
\centering
\includegraphics[trim=1.5cm 4cm 5.3cm 5cm, clip=true, width=\columnwidth]{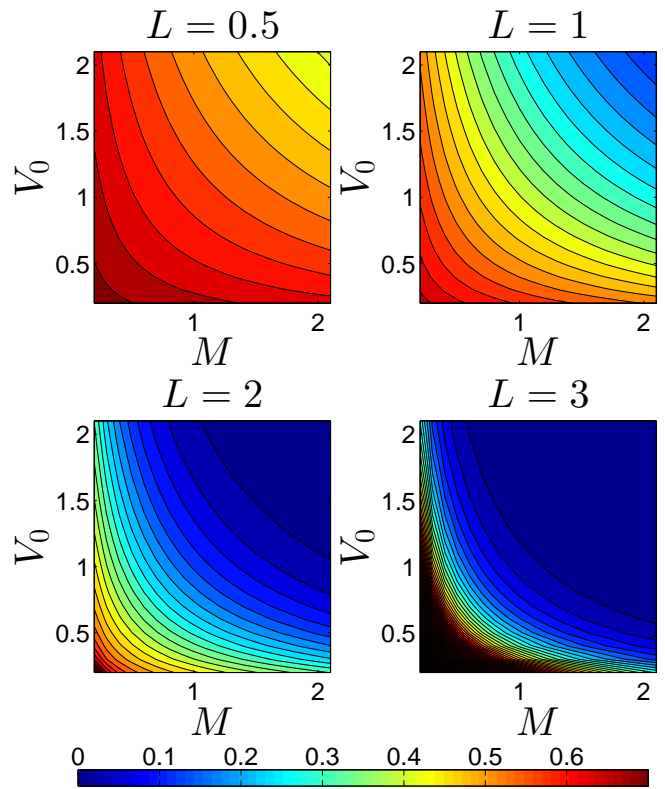}
\caption{The ratio between the energy differences of the first excited state and the ground state to the second excited state and the first ${E_1 - E_0 \over E_2 - E_1}$. When the ratio is much smaller than 1 the second excited state and higher states can be neglected, so the two level approximation is valid. One can see that this occur roughly when $ M V_0 L^2 > \hbar$
}
\label{ratio}
\end{figure}

The spectrum of the mode, for a single unit cell, is obtained by solving the Schr\"{o}dinger equation for the Hamiltonian 
\begin{equation} \label{h1}
H = {P^2 \over 2 M} + V(A),
\end{equation}
where $P$ is the momentum conjugate to $A$ and $ V(A)$ is the double well potential. We use the two level approximation, keeping only the symmetric ground state $\ket{0}$, and the anti symmetric first exited state $\ket{1}$ with excitation energy $ \Gamma$. The range of validity of this approximation is illustrated in Fig \ref{ratio}.
The dispersion of the mode is obtained by considering the coupling to neighboring unit cells, specifically the energy cost $J_{i,j}$ of two unit cells being in different wells. Thus the Hamiltonian is given by
\begin{equation} \label{h2}
H = -{\Gamma \over 2} \sum_i \sigma_x(i) - \sum_{i,j} J_{i,j}\sigma_z (i) \sigma_z (j)
\end{equation}
where $\sigma_{x}(i)$ is the Pauli on site $i$ having the states $\ket{0}$ and $\ket{1} $ as eigenstates with eigenvalues 1 and -1 respectively. 

The Hamiltonian (\ref{h2}), known as the Quantum Ising model, describes a quantum phase transition when $ J \sim \Gamma $. It was widely studied in various dimensionalities and lattices, using several methods\cite{suzuki2013}. Let us just give examples of two results. In one dimension and nearest neighbor coupling $J_{i,j} = J \delta_{j,i\pm1}$, the system can be solved exactly, using the Jordan-Wigner transformation, and the spectrum is given by \cite{suzuki2013}
\begin{equation} \label{w1}
\omega_\mathbf{q} = \Gamma \sqrt{1+ (J/\Gamma)^2 - 2 (J/ \Gamma) \cos q}.
\end{equation}
A more naive calculation, involving the more physical degree of freedom $\sigma_z$ can be done by using a mean field approximation $\sigma_x(i) \sim \langle \sigma_x \rangle \sim 1 $. The resulting frequencies are given by
\begin{equation} \label{w2}
\omega_\mathbf{q} =  \Gamma \sqrt{1 - \langle \sigma_x \rangle J_\mathbf{q} / \Gamma}
\end{equation}
where $J_\mathbf{q} = \sum_j J_{0,j} e^{i \mathbf{R_j} \mathbf{q} } $ is the Fourier transform of the coupling. The critical behavior can be seen by expanding $ {J \over \Gamma} $, in (\ref{w1}) or (\ref{w2}), around $M = M_c$. In both cases, we obtain Eq. (\ref{w0}) with $\omega_s = \Gamma \left( M \left.{ \partial ( J / \Gamma) \over \partial M}\right|_{ M = M_c} \right)^{z \nu}$. In one dimension, Eq. (\ref{w1}), the critical exponent is given by $z \nu = 1$, while the mean field calculation, Eq. (\ref{w2}), yields $z \nu = 1/2$. This also means that even for mean field theory the anomalous isotope effect should be expected as we can see from Eq. (\ref{alphc}): the enhancement of $T_c$ for heavier isotope.

A physical system where this formalism is very likely to be relevant is STO. By specifying the model and connecting the microscopic parameters to observable quantities, we facilitate the search for additional materials with similar phenomena. Our specific predictions regarding the relation of a structural phase transition and superconductivity can focus this search considerably.

In conclusion, we have identified an unusual contribution to the superconducting isotope exponent coming from a proximity to a QCP. This phenomenon can be a distinct experimental signature for the connection between superconductivity and the QCP. Near the QCP this contribution dominates the isotope effect when the critical exponent of the QCP is smaller than one. The relation we derived, Eq. (\ref{alphc}), can be used to quantitatively relate the isotope exponent to the critical exponent. This can be highly useful when there is some experimental data for both quantities.

{\em Acknowledgments}  We are grateful to C. Triola,  S. Pershoguba, J. Edge, U. Aschauer, N. A. Spaldin.
this work was supported by US DOE BES E304. The work of YK was supported by ERC DM 321031 and VR. YK acknowledges the hospitality of LANL.

\end{document}